\documentclass[aip, rsi, amsmath,amssymb,
 reprint,%
floatfix
]{revtex4-1}

\usepackage{graphicx}
\usepackage{dcolumn}
\usepackage{bm}

\usepackage[utf8]{inputenc}
\usepackage[T1]{fontenc}
\usepackage{mathptmx}
\usepackage{etoolbox}
\usepackage{hyperref}

\makeatletter
\def\@email#1#2{%
 \endgroup
 \patchcmd{\titleblock@produce}
  {\frontmatter@RRAPformat}
  {\frontmatter@RRAPformat{\produce@RRAP{*#1\href{mailto:#2}{#2}}}\frontmatter@RRAPformat}
  {}{}
}%
\makeatother
\begin{document}

\preprint{AIP/123-QED}

\title[Reconfigurable Bitter-Type Electromagnet for Zeeman Slowing]{Reconfigurable Bitter-Type Electromagnet for Zeeman Slowing}

\author{Emma G. Hataway}
\author{Emma K. Falk}
\altaffiliation[Now at ]{Department of Physics, University of California, Santa Barbara}
\author{Kaia E. O'Neill}
\altaffiliation[Now at ]{Department of Physics, University of California, Davis}
\author{Morgan P. Berghof}
\author{Ben A. Olsen}%
 \homepage{https://olsenlab.science}
\affiliation{ 
Lewis \& Clark College, Portland, OR 97219, USA
}%
\email{bolsen@lclark.edu}

\date{\today}

\begin{abstract}
Many Zeeman slower magnets are geometrically captured by a vacuum chamber, preventing  modification or repair without breaking vacuum.
We describe a Bitter-type electromagnet coil design that can be easily disassembled, reconfigured, repaired, and replaced without disturbing the vacuum system.
Our coil, designed to slow lithium atoms, produces a near-ideal field profile with a single DC power supply.
With a resistance of 5.7(9)~m$\Omega$, an inductance of 8.8(2)~$\mu$H, and switching times as low as $\sim100$~$\mu$s, the coil compares favorably to other designs, and can be disassembled, modified, and reassembled repeatedly without loss of performance.
With forced-air cooling, the coil experiences  moderate heating.
This coil design offers greater flexibility than traditional electromagnet designs, and it can retrofitted onto existing UHV chambers.
\end{abstract}

\maketitle

\section{\label{sec:intro}Introduction}

One of the most common techniques for slowing beams of atoms and molecules is the Zeeman Slower (ZS) \cite{phillipsLaserDecelerationAtomic1982}.
Using a spatially-varying magnetic field, the Zeeman shift locally counteracts the Doppler shift due to the particles' motion along the beam.
These counteracting shifts allow light from a single-frequency laser to be absorbed by atoms at many positions and velocities in the ZS \cite{firminoProcessStoppingAtoms1990}.
The resulting slowing can reduce the velocity enough to allow capture in a magneto-optical trap (MOT)\cite{raabTrappingNeutralSodium1987}, enabling further laser cooling.

A core ingredient for this process is the ZS magnet, which produces the inhomogeneous field.
Most common designs employ several segments of wire, wrapped around a central tube with varying numbers of turns \cite{phillipsLaserDecelerationAtomic1982,firminoProcessStoppingAtoms1990,martiTwoelementZeemanSlower2010,dedmanOptimumDesignConstruction2004,ohayonInvestigationDifferentMagnetic2015a}. 
The current in each segment is typically controlled by an independent current supply to yield the target field distribution.
These wire-wound ZS coils typically have large resistance and self-inductance, along with slow switching times.
One wire-wound design uses a single wire with variable pitch spacing, reducing the resistance, self-inductance, and overall complexity of the ZS\cite{bellSlowAtomSource2010}.
Our group developed another ZS coil design made from stacks of conducting layers and spacers of different thickness; this Bitter-type coil has faster switching times compared to similar wire-wound designs, and is driven by a single power supply \cite{koiralaEfficientWatercooledBittertype2026}.

One drawback of these current-carrying ZS coil designs is their power dissipation.
Even coils with lower resistance draw $\sim 100$~W of power, leading to heating that can necessitate water cooling.
To overcome this challenge, many groups have employed permanent magnet-based ZS designs.
Some designs employ magnets held in place with 3D-printed forms \cite{parsagianDesigningBuildingPermanent2015}, or CNC-milled holders \cite{liIntegratedHighfluxCold2023}, some have adjustable positions near the atomic beam \cite{hillZeemanSlowersStrontium2014,yuZeemanSlowingGroupIII2022}, while others use Halbach arrays with varying spacing \cite{cheineyZeemanSlowerDesign2011,aliDetailedStudyTransverse2017,wodeyRobustHighfluxSource2021a,marin-bujedoPermanentmagnetZeemanSlower2026}, self-assembled arrays of magnets \cite{lebedevSelfassembledZeemanSlower2014}, or ring-shaped magnets of varying shape \cite{wangLongitudinalZeemanSlower2015}.
While these ZS magnets dissipate no power, their fields cannot be switched off, and can potentially influence later stages of laser cooling.
One hybrid design combines the field switching of a coil with a set of permanent magnets in the highest-field region of the ZS\cite{garwoodHybridZeemanSlower2022}.

Another major drawback shared by many wire-wound and permanent-magnet designs is that they are captured by the Ultra-High Vacuum (UHV) chamber containing the atomic beam.
To reduce power requirements and fringe fields, ZS magnets typically have the smallest inner diameter possible, which makes them impossible to remove once the UHV chamber is assembled.
For such ZS designs, modifications or repairs to the coil are much more challenging, or may even require breaking vacuum, which can be very costly for laser-cooling experiments.
UHV chamber bakeout temperatures can also be limited by the ZS magnets because of the materials commonly used to construct the magnets.

To overcome the geometric capture issue, we set out to design an electromagnet that could be reconfigured and released from the chamber.
Bitter-type electromagnets \cite{bitterDesignPowerfulElectromagnets1936, bitterDesignPowerfulElectromagnets1939, bitterWaterCooledMagnets1962}, made from stacked layers of conducting material rather than wound wires, are well-suited to this reconfigurability.
In this article, we describe the design of a split-layer Bitter-type electromagnet.
We characterize the magnetic field distribution, the coil's electromagnetic properties, and its thermal performance.
We discuss the limitations of this new design and possibilities for overcoming them in future coils.

\section{\label{sec:design} Coil Design} 

\begin{figure*}[ht]
    \centering
    \includegraphics[width=17cm]{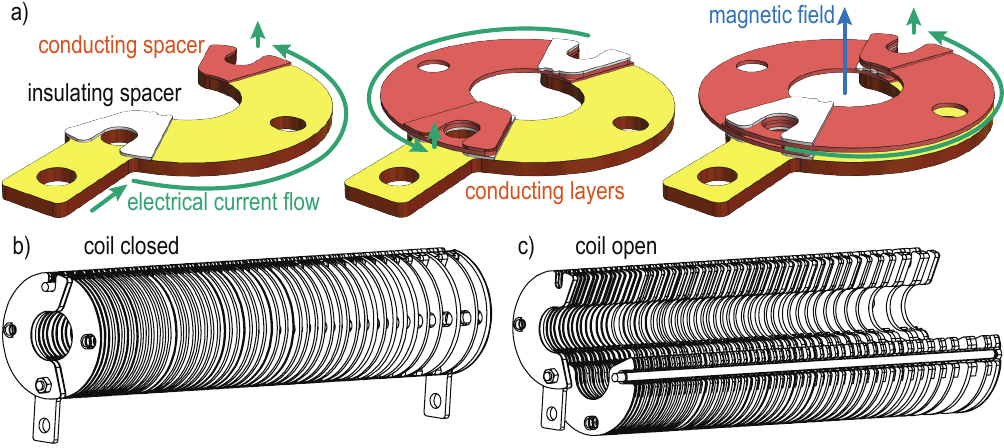}
    \caption{Configuration of the coil. In a), the current (shown in green) flows azimuthally in each layer, then axially in each conducting spacer, leading to a net magnetic field (shown in blue) along the axis of the coil, with local strength set by the layer and spacer thicknesses. Insulating layers of identical thickness lie opposite the conducting spacers between each pair of layers. In b), the complete set of layers and spacers form a solenoid-like shape when closed, held together by tensile forces from a pair of threaded rods and nuts. In c), with the threaded rods loosened, the upper rod can be removed from the slots in the layers, so that every other layer can rotate to open the coil.}
    \label{fig:cad_rbzs}
\end{figure*}

We modified a previous Bitter-type coil design \cite{koiralaEfficientWatercooledBittertype2026} by splitting each conducting layer into two half-layers, separated by insulating or conducting spacers (See Fig.~\ref{fig:cad_rbzs}).
Similar to the previous design, we began with a solenoid-like stack of many layers and spacers of identical thicknesses, then iteratively changed the thicknesses of individual layers and spacers to create the desired magnetic field profile.
We restricted each piece to stock thicknesses available commercially in 101 alloy Oxygen-Free High-Conductivity (OFHC) copper.
Using the \textsc{radia} package for magnetostatics \cite{chubarThreedimensionalMagnetostaticsComputer1998} in \textsc{mathematica}, we computed simulations of the field profile, and compared them to the ideal profile as we modified layer thicknesses.

As seen in Fig.~\ref{fig:cad_rbzs}a, electric current flows azimuthally in each half-layer and axially in each conducting spacer. 
Since the thickness of each layer and spacer is small compared to its radius, the overall coil is roughly solenoidal, leading to a primarily axial magnetic field.
The field due to the axial current flow in nearby spacers roughly cancels at the coil center (the location of the atomic beam).

In normal operation, slots and holes in the even and odd coil layers will align, so that the rough profile of the coil is cylindrical, as seen in Fig.~\ref{fig:cad_rbzs}b.
However, with an upper guide rod removed, all the even coil half-layers can be rotated together relative to the odd half-layers, opening the coil and allowing it to be removed from the UHV chamber, as seen in Fig.~\ref{fig:cad_rbzs}c.
After repair or modification, the coil can be easily reassembled by lining up the slots in each layer and reattaching the guide rod.
The contact between each half-layer and adjacent spacers has fairly large area, so remaking electrical contact should be fairly robust.

With the exception of the end-pieces with attachment points for input and output current wires, all of the half-layers have the same shape.
Similarly, the conducting spacers all have the same shape, as do all the insulating spacers.
This makes modifications to the field profile fairly straightforward, as parts with slightly different thicknesses can be readily swapped in and out.

\subsection{\label{ssec:profile}Magnetic field profile}

Similar to some previous lithium ZS designs\cite{koiralaEfficientWatercooledBittertype2026,mitraExploringAttractivelyInteracting2018, garwoodHybridZeemanSlower2022}, we aimed to create a decreasing-field slower, with a magnetic field profile along the beam axis $\hat z$ given by
\begin{align}
    B(z)=B_\text{bias}+B_0\sqrt{1-\frac{z}{l}}\,,
    \label{eq:Bprofile}
\end{align}
where $B_\text{bias}=h|\delta|/\mu_B$ is determined by the Planck constant $h$, the Bohr magneton $\mu_B$, and the laser detuning $\delta$, $B_0=h v_p/\lambda \mu_B$ is determined by the target velocity $v_p$ of atoms to be slowed and the wavelength $\lambda$ of the laser, and $l$ is the overall length of the ZS. 
In these designs, the field of the ZS coil combines with the field of the MOT coils to produce this desired profile, so that the slowed atoms exiting the ZS are already within the capture region of the MOT.

After several rounds of iteration, we converged on a design with 106 `C'-shaped layers separated by 105 spacer pairs (one conducting, and one insulating).
The coil has length $l=320$~mm, $B_0=100$~mT, and $B_\text{bias}=-26$~mT.
The design field profile matched the ideal profile of Eq.~\ref{eq:Bprofile} to $\sim 1\%$ in our \textsc{radia} simulations.
We verified this simulated profile using a model of the coil produced in \textsc{comsol}, a finite-element simulation package.
In \textsc{comsol}, we also tested the influence of the 316 stainless steel UHV chamber, and found that it did not have a significant effect on the field profile along the ZS axis.

\begin{figure}[ht]
    \centering
    \includegraphics[width=8.5cm]{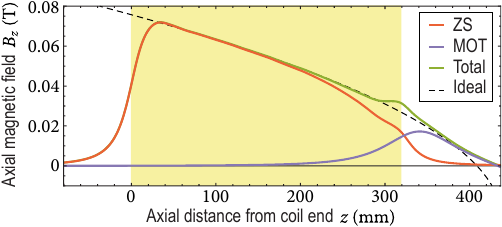}
    \caption{Simulated axial magnetic field profile. The field due to the ZS coil for a current of 261~A is shown in red, and the field due to a pair of MOT coils is shown in purple. The inside of the ZS coil is denoted by the yellow region, with one end 114~mm from the center of the MOT coils. The total field, shown in green, is close to the ideal profile (dashed black), except at the ends of the ZS coil and the center of the MOT coils.}
    \label{fig:fieldprofile}
\end{figure}

\subsection{\label{ssec:sensitivity}Parameter sensitivity}

As a way to test which design parameters have the greatest influence on the field profile, we carried out a series of \textsc{python} simulations of the coils, again using \textsc{radia} models of the layer geometry.
We calculated the fractional change of the magnetic field numerically integrated along the center of the coil while changing various parameters:
\begin{align}
\delta B= \frac{1}{n}\sum_{z}\frac{\left[B(z)_\text{ideal}-B(z)\right] ^2}{ B(z)_\text{ideal}^2}.
 \label{eq:sensitivity}
\end{align}
\begin{figure}
    \centering
    \includegraphics[width=8.5cm]{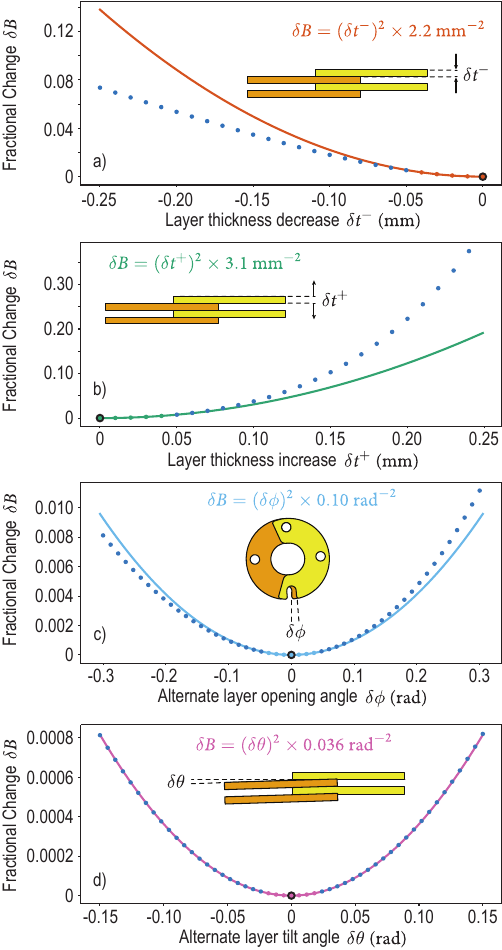}
    \caption{Sensitivity analysis for four deviations from ideal geometry. In each case, the ideal geometry is plotted with a black open circle, and coils with perturbed geometry are shown with colored points. A small number of points (indicated in a different color) near ideal are fit with a parabola, and the quadratic coefficient is shown in each graph. In a), the thickness of each layer decreases by $\delta t^-$, where in b), the thickness of each layer increases by $\delta t^+$. In c), the opening angle (or the azimuthal angle of each even half-layer) $\delta \phi$ of the coil is varied. In d), the tilt angle of each even half-layer $\delta \theta$ is varied. The largest deviation $\delta B$ comes from increasing layer thickness.}
    \label{fig:tolerance}
\end{figure}

We computed the field at  $n=106$ evenly-spaced locations along the $z$-axis of the coil from $z=-60$~mm to $z=420$~mm (see Fig.~\ref{fig:fieldprofile}) to calculate $\delta B$ as a figure of merit for each set of coil parameters.
To find the sensitivity of the field distribution to some expected variations (layer thickness $t$, coil opening angle $\phi$, and inter-layer tilt angle $\theta$), we changed one parameter $\xi$ of the simulated coil over some range, and computed $\delta B(\xi)$ for each geometry.
In all cases, $\delta B(\xi)\propto \xi^2$, so we then fit the resulting values with a second-order polynomial in $\xi$ near its optimum value, as seen in Fig.~\ref{fig:tolerance}.

We found that the field profile is most sensitive to variations in the layer thicknesses $t$, with thicker-than-ideal layers causing greater change, $\delta B=(\delta t^+)^2\times3.1$~mm$^{-2}$, than thinner-than-ideal layers, $\delta B=(\delta t^-)^2\times2.2$~mm$^{-2}$.
Based on manufacturer tolerances of $\delta t=\pm0.006"=\pm0.15$~mm for the most common layer thickness\footnote{\href{https://www.mcmaster.com/89675K765}{www.mcmaster.com/89675K765}}, we predict $\delta B = 0.070, 0.050$.
These estimates are based on each layer thickness $t$ varying by the same amount, rather than randomly, which would lead to an overall smaller deviation from ideal.
Since many half-layers are cut from the same piece of stock in our construction method, as described below, a correlated variation in thickness seems reasonable for our coil.

The coil opening angle $\phi$ yielded a fractional change $\delta B=( \delta\phi)^2\times0.10$~rad$^{-2}$, and the even-layer tilt produced $\delta B=(\delta \theta)^2\times0.036$~rad$^{-2}$.
To produce a comparable $\delta B$ to that produced by the layer thickness variation, these angles would need to be greater than $0.7$~rad.
For values of $\delta \phi$ and $\delta \theta$ expected for coil construction, we can treat these deviations as negligible.
Variations in $\phi$ also lead to small changes in the contact resistance between layers, as we will discuss in Sec.~\ref{sec:electromagnetic}.

\subsection{\label{ssec:construction}Construction}

To create the layers and spacers, we sent sheets of stock OFHC copper and fiberglass to a commercial machine shop, where they milled the fiberglass parts and laser-cut the copper pieces.
Many university machine shops have the capability to do this sort of cutting.
This method ensured that the faces of the layers remained flat and parallel.
We sanded each of the cut copper pieces with 2000 grit sandpaper, submerged them in 70~$^\circ$C acetic acid for roughly 3~min, then rinsed them in distilled water and dried them. 
Next, we silver-plated the contact surfaces, as described in detail in Sec.~\ref{ssec:surface}.

We fastened each spacer to one of the adjoining copper layers using epoxy\footnote{JB Weld 2-part epoxy} around the spacer perimeter, making sure to avoid any between conducting pieces.
We stacked the layer/spacer assemblies using a pair of threaded rods insulated by plastic tubes as guides (See Fig.~\ref{fig:cad_rbzs}b). 
Each layer aligns so one of the rods passes through a hole, and the other passes through an open slot.
The rod passing through the slots can be easily slid radially outward, allowing the layers to rotate about the rod passing through the holes (see Fig.~\ref{fig:cad_rbzs}c).
Once we had stacked all the layers, we tightened nuts on the ends of the threaded rods to a torque of 6~N\,m. 
We empirically determined this torque by measuring the resistance of the coil as we increased the tightening torque: it plateaued above 5~N\,m, and the fasteners began to deform at higher torques.

\section{\label{sec:electromagnetic}Electromagnetic Properties}
\subsection{Magnetic field distribution}

\begin{figure}[ht]
    \centering
    \includegraphics[width=8.5cm]{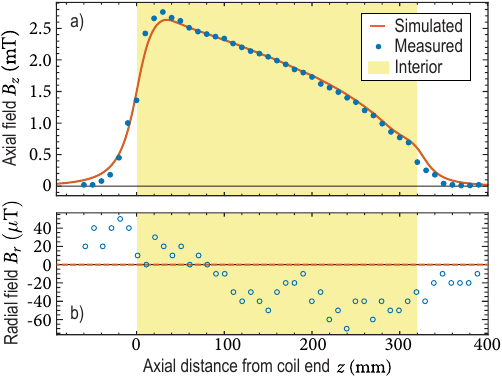}
    \caption{Measured axial (filled circle) and radial (open circle) components of the magnetic field, $B_z, B_r$, along the central axis of the coil for a current of $10$~A. Except near the edges of the coil, the axial field profile matches simulations well. While our simulations predict a vanishing radial field, we measured a small component near the resolution of our radial field probe (10~$\mu$T).}
    \label{fig:measuredbfield}
\end{figure}

Using an axial magnetic field probe\footnote{AlphaLab Inc. GM2 with high-stability axial probe} centered in the bore of the coil and insulated with a plastic tube, we measured the magnetic field profile $B_z(z)$ for a current $I=10$~A (see Fig.~\ref{fig:measuredbfield}a).
The field profile matches our simulations well, except for some deviation near the edges of the coil.
We also measured the radial magnetic field profile $B_r(z)$ with a transverse-field probe \footnote{AlphaLab Inc. GM2 with universal transverse-field probe} (see Fig.~\ref{fig:measuredbfield}b).
While our simulations predict a purely axial field, we found a small radial component near the measurement resolution of our probe.
\subsection{\label{ssec:impedance}Electrical impedance and mutual inductance}

\begin{figure}[ht]
    \centering
    \includegraphics[width=8.5cm]{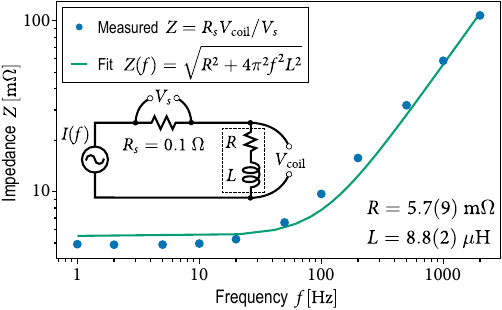}
    \caption{Impedance $Z$ of the ZS coil as the applied AC current frequency $f$ is varied. A schematic of the measurement circuit is shown in the inset. We measured the applied current with a sense resistor $R_s$, yielding $V_s=IR_s$. The voltage of the coil is $V_\text{coil}=IZ$, so we can compute the impedance $Z(f)=V_\text{coil}(f)R_s/V_s(f)$. We fit the data with a lumped $RL$ circuit model, which yielded best-fit values $R=5.7(9)$~m$\Omega, L=8.8(2)~\mu$H.}
    \label{fig:impedance}
\end{figure}

To measure the resistance of the coil, we applied DC a current of $200$~A and measured the voltage drop. We found $V_\text{coil}=1.83(2)$~V, and $V_\text{leads}=0.77(2)$~V with the coil removed from the circuit, yielding a resistance $R=5.3(2)~$m$\Omega$.
We also measured the AC impedance $Z(f)$ of the coil using the circuit depicted in the inset of Fig.~\ref{fig:impedance}---an AC current source drove a sense resistor $R_s$ in series with the ZS coil (which we modeled as a lumped $RL$ circuit).
As we varied the frequency $f$ of the AC current $I(f)$, we measured $V_s(f)=I(f)R_s$ and $V_\text{coil}=I(f)Z(f)$. 
The impedance of the ZS coil was then $Z(f)=R_s V_\text{coil}(f)/V_s(f)$.
We varied $f$ over three orders of magnitude, and fit the the resulting data with a lumped $RL$ model with $Z(f)=\sqrt{R^2+4\pi^2f^2L^2}$, shown in Fig.~\ref{fig:impedance}.
The best-fit resistance $R=5.7(9)$~m$\Omega$ is consistent with the DC value.
The best-fit inductance, $L=8.8(2)~\mu$H, is about half as large as a previous coil design with similar size, but more layers/turns\cite{koiralaEfficientWatercooledBittertype2026}, as expected.

Using the same circuit to drive the ZS coil, we measured the voltage across a pair of MOT coils located nearby (See Fig.~\ref{fig:fieldprofile}). With an oscillating current $I(f)=I_0 \sin(2\pi f t)$ in the ZS coil, a voltage $V_\text{MOT}(f)=-M \frac{d I(f)}{dt}=-M 2 \pi f I_0 \cos(2 \pi f t)=-V_0 \cos(2\pi f t)$, so $M=R_s V_0/2\pi f V_s$.
Applying an AC current at $f=1$~kHz, we measured the voltage across the pair of MOT coils, in both anti-Helmholtz and Helmholtz configurations.
In Helmholtz configuration, we found $M_\text{H}=0.02(2)~\mu$H. 
Based on the arrangement of the MOT coils, we would expect a vanishing $M_\text{H}$ for perfectly symmetric alignment.
For anti-Helmholtz configuration, we measured $M_\text{aH}=0.11(2)~\mu$H.
This value, combined with the field switching profiles measured below, predicts a small induced emf of $2.2$~V in the MOT coils as the ZS coil is shut off.
\subsection{\label{ssec:switching}Field switching}

\begin{figure}[ht]
    \centering
    \includegraphics[width=8.5cm]{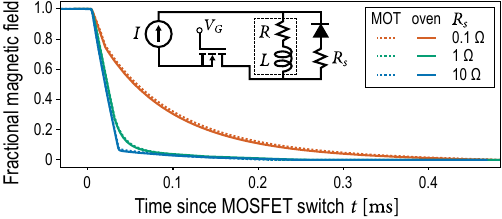}
    \caption{Measurements of the field switching time. An initial current of $I=200$~A flowed through the coil and a MOSFET, until at $t=0$, we switched the gate voltage $V_G$ of the MOSFET low. Current then flowed through the coil, a shunt resistor $R_s$, and a protection diode. We measured the induced emf of a small pickup loop with several turns (located either in the MOT end of the coil or the oven end), which we integrated to find $B(t)$. We see an initial fast linear decay in the field, followed by non-exponential decay with characteristic time roughly given by $L/(R+R_s)$. In all cases, the field decayed to zero within $500~\mu$s.}
    \label{fig:switching}
\end{figure}

We measured the field switching characteristic time for the ZS coil using the circuit depicted in the inset of Fig.~\ref{fig:switching}.
Current from a DC power supply was controlled with a MOSFET---when the gate voltage $V_G$ was high, it conducted, and the DC current passed through the ZS coil and the MOSFET.
A protection diode prevented current from flowing from the supply through a shunt resistor $R_s$.
When we suddenly brought $V_G$ low, the MOSFET stopped conducting, and current flowed through the shunt resistor and protection diode.
We measured the axial field using a small loop of wire with several turns, yielding a voltage $V\propto dB/dt$, which we integrated to compute $B(t)$, shown in Fig.~\ref{fig:switching}.
At both ends of the coil (the oven/high-field end, and the low-field/MOT end), the field decay was linear at first, followed by a non-exponential decay, with characteristic time depending on the value of $R_s$.
In all cases, the field decayed to zero within $500~\mu$s.
Compared to a coil with similar dimensions but roughly double $L$\cite{koiralaEfficientWatercooledBittertype2026}, measured using the same circuit, this coil exhibits field switching approximately twice as fast.

\subsection{\label{ssec:surface}Interface electrical resistance}

This coil design has a large number of interfaces between conducting surfaces---110 flat face pairs held together by axial compressive forces.
During anticipated operation of the coil, it will occasionally be loosened, opened, modified, and re-tightened, so the interfaces will be repeatedly exposed to air.
Copper has a tendency to corrode when exposed to air\cite{riceAtmosphericCorrosionCopper1981}, and in turn, undergoes an increase in surface resistivity.
Without a surface treatment, the resistance of the coil will increase over time due to the corrosion of the copper\cite{braunovicElectricalContactsFundamentals2017}.
A common solution to this issue is plating the exposed contact surfaces with a thin layer of another metal to maintain the lowest surface resistivity\cite{khayamReducingElectricalContact2013}.

\begin{figure}[ht]
    \centering
    \includegraphics[width = 8.5cm]{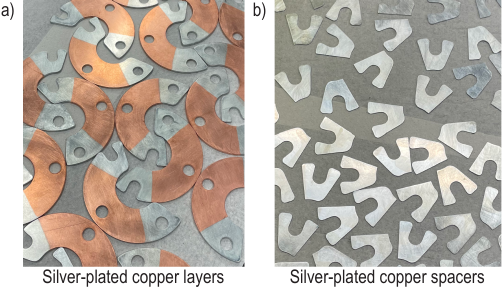}
    \caption{a) Photo of copper layers after plating connecting surfaces with silver. b) Photo of copper spacers after plating with silver. This plating led to the least increase in coil resistance after disassembling and reassembling the coil.}
    \label{fig:silver_plating}
\end{figure}

To test the effects of various surface treatments, we measured the resistance of several stacks of copper discs with similar thickness to the components of the coil.
We intentionally corroded the surfaces of one group, polished one group, plated one group with nickel, and plated one group with silver. 
Nickel plating is commonly used for protecting metals, as it reduces surface corrosion\cite{sadiku-agboolaInfluenceOperationParameters2011}. 
While the resistivity of silver is lower than that of copper, silver also tarnishes when exposed to air, forming Ag$_2$S and AgCl\cite{graedelCorrosionMechanismsSilver1992}, with a rate mostly independent of atmospheric humidity, but dependent on sulfur concentration \cite{riceAtmosphericCorrosionCopper1981}.
At the plating interface, Cu(Ag) alloys can have much lower resistance than other alloys \cite{strehleElectricalPropertiesElectroplated2009}.
While gold plating can lead to less corrosion and more durable electrical contacts, gold-plated copper leads to diffusion of copper into the gold, leading to higher resistivity in the alloy \cite{pucicDiffusionCopperGold1993}.

Before applying surface treatments, we sanded each copper piece with 1200-grit sandpaper, washed in glacial acetic acid at $70~^\circ\text{C}$ for 3 minutes, then rinsed with distilled water before drying.
For the corrosion group, we immersed the discs in boiling water for 10 minutes, and left them exposed to laboratory air.
For pieces in the polishing treatment group,  we sanded, dried, and hand-polished the surfaces using a commercial polish\footnote{Hagerty 100, a Kaolin Clay-based polish}.
For the nickel group, we bath electroplated nickel onto both sides of the copper test pieces using a commercial solution\footnote{Gold Plating Services---Bright Nickel Plating Solution}, a nickel electrode, and a laboratory DC power supply: 15~s at $\sim 3$~V on both the front and back of the piece.
Similarly, we electroplated silver using a commercial solution\footnote{Gold Plating Services---Bright Silver NC Bath solution}, a stainless steel anode, and DC current for $10$~s per side at $\sim 2$~V.

After applying the surface treatments, we clamped the discs in each group together (we estimate a compressive force of 3750~N), and found their resistance by applying a DC current and measuring the resulting voltage drop. 
We took resistance measurements over the span of one week to determine how the total resistance of the stacks changed. 
In between measurements, the pieces were unclamped and exposed to air.
During initial testing, we also found that touching the pieces without wearing gloves led to a significant increase in resistance, likely due to skin oils on the metal surfaces. 

\begin{table}[hbt]
    \centering
    \begin{tabular}{ | c | c | c | }
    \hline
    Surface treatment & $R_i$~(m$\Omega$) & $R_\text{week}$~(m$\Omega$)\\
    \hline
    Oxidized & 0.14(1) & 0.42(1)\\
    Polished & 0.14(1) & 0.15(1)\\
    Nickel & 0.22(1) & 0.51(1)\\
    Silver & 0.10(1) & 0.10(1)\\
    \hline
    \end{tabular}
    \caption{Electrical resistance of stacks of copper layers with various surface treatments. We measured the stacks immediately after applying the surface treatments to find the initial resistances $R_i$ (applying a DC current of $1.00$~A and measuring the voltage drop). After one week with the coil stacks disassembled and exposed to laboratory air, we re-measured them to find $R_\text{week}$. The oxidized and nickel-plated treatments had significantly increased $R$, while the polished and silver-plated treatments showed no significant increase.}
    \label{tab:surface_res}
\end{table}

We found that the initial surface treatment changed the stack resistance: the nickel plating increased $R$, and silver plating decreased $R$, as expected given the relative resistivities of Cu, Ni, and Ag.
After one week, the oxidized and nickel-plated groups' $R$ increased significantly, while the polished and silver-plated groups' $R$ did not significantly change. 
Based on these results, we opted to silver-plate the contact surfaces of the Zeeman slower coil, as seen in Fig.~\ref{fig:silver_plating}.

\section{\label{sec:thermal}Thermal Properties}
\subsection{\label{ssec:thermalCalc}Simulations}

\begin{figure}[ht]
    \centering
    \includegraphics[width=8.5cm]{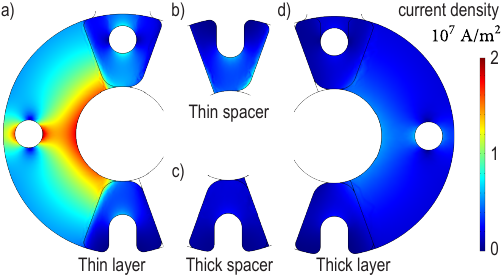}
    \caption{Finite-element simulations of the current density in various parts of the coil. In a), the thin layer shows the highest local current density, reaching nearly $2\times10^7$~A/m$^2$ near the inner radial edge. In b) and c), the thin and thick spacers show lower current densities, since current primarily flows axially. In d), the thick layer reaches only about $6\times10^6$~A/m$^2$.}
    \label{fig:current_density}
\end{figure}

Since the cross-section of the coil changes along the path of the current, we used finite element analysis in {\scshape comsol} to simulate the local current density (see Fig.~\ref{fig:current_density}) for a total applied current of 200~A.
As expected, the thinner layers showed a higher average current density, with the greatest magnitude near the inner radial edge, and near the circular hole in the layers.

\begin{figure}[ht]
    \centering
    \includegraphics[width=8.5cm]{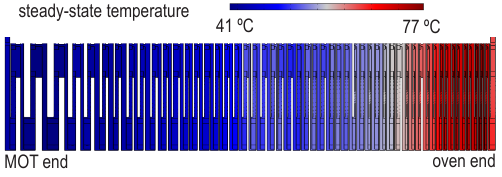}
    \caption{Finite-element simulation of the temperature distribution in the coil after an experimental cycle (side view). We see that the MOT end, with overall thicker copper, heats less due to the lower average current density, while the oven end heats more.}
    \label{fig:temp_dist}
\end{figure}

Taking into account this inhomogeneous current density, we computed the resulting temperature in the coil after a typical experimental cycle time of 1~min with 200~A of current, as seen in Fig.~\ref{fig:temp_dist}.
We see that, near the MOT end of the coil, where the layers and spacers are thicker (with lower current densities), the resulting temperature is lower, and reaches a value near $41~^\circ$C.
Near the oven end of the coil, the predicted temperature rises to $77~^\circ$C.

\subsection{\label{ssec:heating}Heating measurements}

\begin{figure}[ht]
    \centering
    \includegraphics[width=8.5cm]{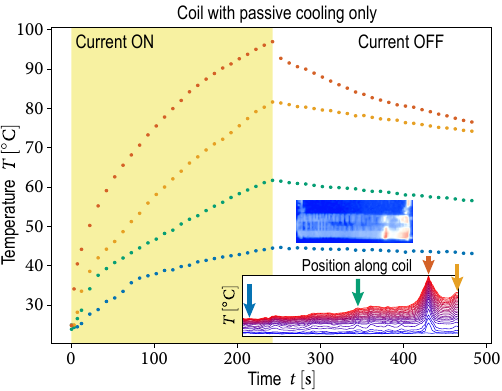}
    \caption{Heating of the coil with passive convective cooling only. For the first 240~s, we applied 200~A of current and periodically measured the temperature distribution of the coil using a thermal camera (see inset). Averaging perpendicular to the axis of the coil, we computed the profiles at various times shown in the inset. A hot spot in the coil reached the highest temperatures, likely due to a single faulty contact. We show the temperature of the hot spot and three other locations on the coil as functions of time.}
    \label{fig:heating_passive}
\end{figure}

To measure the thermal properties of the ZS coil, we passed $200$~A of DC current through the coil for 240~s (much longer than the expected duration during a typical experimental cycle) and imaged the coil using a thermal camera, as seen in Fig.~\ref{fig:heating_passive}.
We integrated along the vertical axis of the images to measure an average temperature distribution along the axial length of the coil (colored curves in the inset).
In the profiles, we see that the high-field end of the coil reached a higher temperature (similar to our simulations), but a `hot spot' developed in the inner part of the coil, likely due to a single faulty contact between layers and spacers.
We show the temperature evolution of four points along the coil over 240~s of heating and 260~s of passive cooling.

\begin{figure}[ht]
    \centering
    \includegraphics[width=8.5cm]{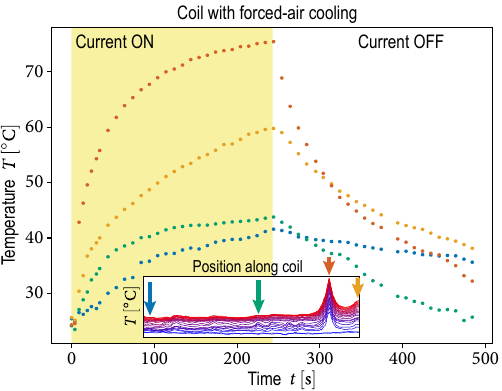}
    \caption{Heating of the coil with forced-air cooling using a fan. For the first 240~s, we applied 200~A of current and periodically measured the temperature distribution of the coil using a thermal camera. Averaging perpendicular to the axis of the coil, we computed the profiles at various times shown in the inset. The hot spot and three other locations on the coil reached lower maximum temperatures than in the passive cooling case, and once the current was turned off, began to decrease much faster.}
    \label{fig:heating_air}
\end{figure}

In another test, we repeated the same heating measurements, but with forced-air cooling (with air speed around 3~m/s) applied to the entire coil, as seen in Fig.~\ref{fig:heating_air}.
The maximum resulting temperatures were significantly lower, and returned to room temperature much more quickly.
Compared to a similar-size coil design with water cooling \cite{koiralaEfficientWatercooledBittertype2026}, this coil experienced roughly four times greater heating, but with significantly simpler construction and setup requirements.

\section{\label{sec:conclusion}Conclusion}

We have described the design, construction, and characterization of a novel, reconfigurable electromagnet coil for Zeeman slowing in cold-atom experiments.
The coil produces a magnetic field profile that matches the ideal slowing profile well, with small resistance and self-inductance.
The field switches off in a time compatible with typical cold-atom loading sequences, and requires only a single laboratory DC power supply.
The coil heats more than similar designs, but with no water cooling.
Due to the constant outer diameter of the coil, a simple outer cooling jacket could be used to limit temperature rise.
Another design modification that could reduce the coil size and heating would be to decrease the inner diameter of the layers near the oven end of the coil.

After silver-plating the contact surfaces, the coil can be disassembled, reconfigured, and reassembled without altering the UHV system it attaches to, and without sacrificing coil performance.
This coil design could also be used to retrofit existing UHV chambers under vacuum whose ZS coils suffer degradation.

\section*{Appendix}

Here we provide a list of materials used in construction of this coil. Since its design is simpler than other ZS coil designs, there are relatively few parts.
\begin{table}[ht]
    \centering
    \begin{tabular}{lcc}
         Material & Quantity & Cost\\ \hline
         101 OFHC Copper sheet ($12"\times12"\times0.04"$) & 3 & \$73.78\\
         101 OFHC Copper sheet ($12"\times12"\times0.08"$)& 1 & \$134.16\\
         101 OFHC Copper sheet ($12"\times12"\times0.125"$)& 1 & \$192.15\\
         G10 Fiberglass sheet ($12"\times36"\times0.04"$) & 1 & \$65.15\\
         G10 Fiberglass sheet  ($12"\times12"\times0.125"$)& 1 & \$185.99\\
         Threaded Rod ($1/4"$--$20\times24"$)& 2 & \$8.10\\
         Polycarbonate Tube ($1/4" \text{~ID}\times 24"$)& 2 & \$10.91\\
         Silver Plating Solution (1~L) & 1 & \$315\\
         Stainless Steel Anode $(2"\times6"\times0.075")$ &  1 & \$8.73\\
         Epoxy (10 packets) & 1 & \$23.74\\
         \hline
         & & \$1,184.28
    \end{tabular}
    \caption{Materials used in the construction of the final design ZS coil, with approximate costs from COTS vendors\footnote{Most stock material was purchased at McMaster-Carr.}. Additional small components, like nuts and washers, add negligible cost.}
    \label{tab:materials}
\end{table}

\begin{acknowledgments}
B.\,A.\,O. acknowledges support from the M.\,J.\,Murdock Charitable Trust, and the National Science Foundation through Grant No.\,PHY-2418777.
\end{acknowledgments}

\section*{Data Availability Statement}

The data that support the findings of this study are openly available in Zenodo at http://doi.org/[doi], reference number [reference number].
An updated version can be accessed at\footnote{\href{https://github.com/olsenlab-science/Claw-ZS}{github.com/olsenlab-science/Claw-ZS}}.

\bibliography{2025_CZS}

@article{aliDetailedStudyTransverse2017,
  title = {Detailed Study of a Transverse Field {{Zeeman}} Slower},
  author = {Ali, D Ben and Badr, T and Br{\'e}zillon, T and Dubessy, R and Perrin, H and Perrin, A},
  year = 2017,
  month = mar,
  journal = {Journal of Physics B: Atomic, Molecular and Optical Physics},
  volume = {50},
  number = {5},
  pages = {055008},
  publisher = {IOP Publishing},
  issn = {0953-4075, 1361-6455},
  doi = {10.1088/1361-6455/aa5a6a},
  urldate = {2025-07-28},
  copyright = {http://iopscience.iop.org/info/page/text-and-data-mining},
  langid = {english}
}

@article{bellSlowAtomSource2010,
  title = {A Slow Atom Source Using a Collimated Effusive Oven and a Single-Layer Variable Pitch Coil {{Zeeman}} Slower},
  author = {Bell, S. C. and Junker, M. and Jasperse, M. and Turner, L. D. and Lin, Y.-J. and Spielman, I. B. and Scholten, R. E.},
  year = 2010,
  month = jan,
  journal = {Review of Scientific Instruments},
  volume = {81},
  number = {1},
  pages = {013105},
  issn = {0034-6748, 1089-7623},
  doi = {10.1063/1.3276712},
  urldate = {2020-06-04},
  langid = {english}
}

@article{bitterDesignPowerfulElectromagnets1936,
  title = {The {{Design}} of {{Powerful Electromagnets Part II}}. {{The Magnetizing Coil}}},
  author = {Bitter, F.},
  year = 1936,
  month = dec,
  journal = {Review of Scientific Instruments},
  volume = {7},
  number = {12},
  pages = {482--488},
  issn = {0034-6748, 1089-7623},
  doi = {10.1063/1.1752068},
  urldate = {2024-06-21},
  langid = {english}
}

@article{bitterDesignPowerfulElectromagnets1939,
  title = {The {{Design}} of {{Powerful Electromagnets Part IV}}. {{The New Magnet Laboratory}} at {{M}}. {{I}}. {{T}}.},
  author = {Bitter, F.},
  year = 1939,
  month = dec,
  journal = {Review of Scientific Instruments},
  volume = {10},
  number = {12},
  pages = {373--381},
  issn = {0034-6748, 1089-7623},
  doi = {10.1063/1.1751470},
  urldate = {2024-05-31},
  langid = {english}
}

@article{bitterWaterCooledMagnets1962,
  title = {Water {{Cooled Magnets}}},
  author = {Bitter, F.},
  year = 1962,
  month = mar,
  journal = {Review of Scientific Instruments},
  volume = {33},
  number = {3},
  pages = {342--349},
  issn = {0034-6748, 1089-7623},
  doi = {10.1063/1.1717838},
  urldate = {2024-05-31},
  langid = {english}
}

@book{braunovicElectricalContactsFundamentals2017,
  title = {Electrical {{Contacts}}: {{Fundamentals}}, {{Applications}} and {{Technology}}},
  shorttitle = {Electrical {{Contacts}}},
  author = {Braunovic, Milenko and Myshkin, Nikolai K. and Konchits, Valery V.},
  year = 2017,
  month = dec,
  publisher = {CRC Press},
  address = {Boca Raton},
  doi = {10.1201/9780849391088},
  isbn = {978-1-315-22219-6}
}

@article{cheineyZeemanSlowerDesign2011,
  title = {A {{Zeeman}} Slower Design with Permanent Magnets in a {{Halbach}} Configuration},
  author = {Cheiney, P. and Carraz, O. and {Bartoszek-Bober}, D. and Faure, S. and Vermersch, F. and Fabre, C. M. and Gattobigio, G. L. and Lahaye, T. and {Gu{\'e}ry-Odelin}, D. and Mathevet, R.},
  year = 2011,
  month = jun,
  journal = {Review of Scientific Instruments},
  volume = {82},
  number = {6},
  pages = {063115},
  issn = {0034-6748, 1089-7623},
  doi = {10.1063/1.3600897},
  urldate = {2025-02-17},
  langid = {english}
}

@article{chubarThreedimensionalMagnetostaticsComputer1998,
  title = {A Three-Dimensional Magnetostatics Computer Code for Insertion Devices},
  author = {Chubar, Oleg and Elleaume, Pascal and Chavanne, Joel},
  year = 1998,
  month = may,
  journal = {Journal of Synchrotron Radiation},
  volume = {5},
  number = {3},
  pages = {481--484},
  issn = {0909-0495},
  doi = {10.1107/S0909049597013502},
  urldate = {2025-08-19},
  langid = {english}
}

@article{dedmanOptimumDesignConstruction2004,
  title = {Optimum Design and Construction of a {{Zeeman}} Slower for Use with a Magneto-Optic Trap},
  author = {Dedman, C. J. and Nes, J. and Hanna, T. M. and Dall, R. G. and Baldwin, K. G. H. and Truscott, A. G.},
  year = 2004,
  month = dec,
  journal = {Review of Scientific Instruments},
  volume = {75},
  number = {12},
  pages = {5136--5142},
  issn = {0034-6748, 1089-7623},
  doi = {10.1063/1.1820524},
  urldate = {2020-06-01},
  langid = {english}
}

@article{firminoProcessStoppingAtoms1990,
  title = {Process of Stopping Atoms with the {{Zeeman}} Tuning Technique with a Single Laser},
  author = {Firmino, M. E. and Faria Leite, C. A. and Zilio, S. C. and Bagnato, V. S.},
  year = 1990,
  month = apr,
  journal = {Physical Review A},
  volume = {41},
  number = {7},
  pages = {4070--4073},
  issn = {1050-2947, 1094-1622},
  doi = {10.1103/PhysRevA.41.4070},
  urldate = {2025-02-15},
  copyright = {http://link.aps.org/licenses/aps-default-license},
  langid = {english}
}

@article{garwoodHybridZeemanSlower2022,
  title = {A Hybrid {{Zeeman}} Slower for Lithium},
  author = {Garwood, Davis and Liu, Liyu and Mongkolkiattichai, Jirayu and Yang, Jin and Schauss, Peter},
  year = 2022,
  month = mar,
  journal = {Review of Scientific Instruments},
  volume = {93},
  number = {3},
  pages = {033202},
  issn = {0034-6748},
  doi = {10.1063/5.0081080},
  urldate = {2024-01-21}
}

@article{graedelCorrosionMechanismsSilver1992,
  title = {Corrosion {{Mechanisms}} for {{Silver Exposed}} to the {{Atmosphere}}},
  author = {Graedel, T. E.},
  year = 1992,
  month = jul,
  journal = {Journal of The Electrochemical Society},
  volume = {139},
  number = {7},
  pages = {1963},
  publisher = {IOP Publishing},
  issn = {1945-7111},
  doi = {10.1149/1.2221162},
  urldate = {2025-09-16},
  langid = {english}
}

@article{hillZeemanSlowersStrontium2014,
  title = {Zeeman {{Slowers}} for {{Strontium}} Based on {{Permanent Magnets}}},
  author = {Hill, Ian R. and Ovchinnikov, Yuri B. and Bridge, Elizabeth M. and Curtis, E. Anne and Gill, Patrick},
  year = 2014,
  month = apr,
  journal = {Journal of Physics B: Atomic, Molecular and Optical Physics},
  volume = {47},
  number = {7},
  eprint = {1402.5271},
  primaryclass = {physics},
  pages = {075006},
  issn = {0953-4075, 1361-6455},
  doi = {10.1088/0953-4075/47/7/075006},
  urldate = {2024-05-31},
  archiveprefix = {arXiv},
  langid = {english}
}

@inproceedings{khayamReducingElectricalContact2013,
  title = {Reducing Electrical Contact Resistance at Highly Loaded Copper Conductor Using Nickel and Silver Coating},
  booktitle = {2013 {{Joint International Conference}} on {{Rural Information}} \& {{Communication Technology}} and {{Electric-Vehicle Technology}} ({{rICT}} \& {{ICeV-T}})},
  author = {Khayam, Umar and Risdiyanto, Agus and {Suwarno}},
  year = 2013,
  month = nov,
  pages = {1--6},
  doi = {10.1109/rICT-ICeVT.2013.6741534},
  urldate = {2025-09-16}
}

@article{koiralaEfficientWatercooledBittertype2026,
  title = {Efficient Water-Cooled {{Bitter-type}} Electromagnet for {{Zeeman}} Slowing in Cold-Atom Experiments},
  author = {Koirala, Rishav and Olsen, Ben A.},
  year = 2026,
  month = mar,
  journal = {Review of Scientific Instruments},
  volume = {97},
  number = {3},
  pages = {033205},
  issn = {0034-6748, 1089-7623},
  doi = {10.1063/5.0309624},
  urldate = {2026-03-27},
  langid = {english}
}

@article{lebedevSelfassembledZeemanSlower2014,
  title = {Self-Assembled {{Zeeman}} Slower Based on Spherical Permanent Magnets},
  author = {Lebedev, V and Weld, D M},
  year = 2014,
  month = aug,
  journal = {Journal of Physics B: Atomic, Molecular and Optical Physics},
  volume = {47},
  number = {15},
  pages = {155003},
  issn = {0953-4075, 1361-6455},
  doi = {10.1088/0953-4075/47/15/155003},
  urldate = {2025-02-17},
  copyright = {http://iopscience.iop.org/info/page/text-and-data-mining},
  langid = {english}
}

@article{liIntegratedHighfluxCold2023,
  title = {An Integrated High-Flux Cold Atomic Beam Source for Strontium},
  author = {Li, Jie and Jia, Zhi-Peng and Liu, Peng and Liu, Xiao-Yong and Wang, De-Zhong and Kong, De-Quan and Li, Su-Peng and Cui, Xing-Yang and Dai, Han-Ning and Chen, Yu-Ao and Pan, Jian-Wei},
  year = 2023,
  month = sep,
  journal = {Review of Scientific Instruments},
  volume = {94},
  number = {9},
  pages = {093202},
  issn = {0034-6748, 1089-7623},
  doi = {10.1063/5.0162128},
  urldate = {2025-09-03},
  langid = {english}
}

@article{marin-bujedoPermanentmagnetZeemanSlower2026,
  title = {A Permanent-Magnet {{Zeeman}} Slower and Magneto-Optical Trap for Calcium Atoms for Ultracold {{Rydberg}} Physics},
  author = {{Marin-Bujedo}, E. and Grondin, J. A. L. and Schiltz, T. and Corbo, T. and Urbain, X. and G{\'e}n{\'e}vriez, M.},
  year = 2026,
  month = mar,
  journal = {Review of Scientific Instruments},
  volume = {97},
  number = {3},
  pages = {033202},
  issn = {0034-6748, 1089-7623},
  doi = {10.1063/5.0314204},
  urldate = {2026-06-03},
  langid = {english}
}

@article{martiTwoelementZeemanSlower2010,
  title = {Two-Element {{Zeeman}} Slower for Rubidium and Lithium},
  author = {Marti, G. Edward and Olf, Ryan and Vogt, Enrico and {\"O}ttl, Anton and {Stamper-Kurn}, Dan M.},
  year = 2010,
  month = apr,
  journal = {Physical Review A},
  volume = {81},
  number = {4},
  pages = {043424},
  issn = {1050-2947, 1094-1622},
  doi = {10.1103/PhysRevA.81.043424},
  urldate = {2025-02-14},
  copyright = {http://link.aps.org/licenses/aps-default-license},
  langid = {english}
}

@phdthesis{mitraExploringAttractivelyInteracting2018,
  title = {Exploring Attractively Interacting Fermions in {{2D}} Using a {{Quantum Gas Microscope}}},
  author = {Mitra, Debayan},
  year = 2018,
  month = nov,
  langid = {english},
  school = {Princeton University}
}

@article{ohayonInvestigationDifferentMagnetic2015a,
  title = {Investigation of Different Magnetic Field Configurations Using an Electrical, Modular {{Zeeman}} Slower},
  author = {Ohayon, Ben and Ron, Guy},
  year = 2015,
  month = oct,
  journal = {Review of Scientific Instruments},
  volume = {86},
  number = {10},
  pages = {103110},
  issn = {0034-6748, 1089-7623},
  doi = {10.1063/1.4934248},
  urldate = {2025-09-16},
  langid = {english}
}

@article{parsagianDesigningBuildingPermanent2015,
  title = {Designing and Building a Permanent Magnet {{Zeeman}} Slower for Calcium Atoms Using a {{3D}} Printer},
  author = {Parsagian, Alexandria and Kleinert, Michaela},
  year = 2015,
  month = oct,
  journal = {American Journal of Physics},
  volume = {83},
  number = {10},
  pages = {892--899},
  issn = {0002-9505, 1943-2909},
  doi = {10.1119/1.4930080},
  urldate = {2025-02-15},
  langid = {english}
}

@article{phillipsLaserDecelerationAtomic1982,
  title = {Laser {{Deceleration}} of an {{Atomic Beam}}},
  author = {Phillips, William D. and Metcalf, Harold},
  year = 1982,
  month = mar,
  journal = {Physical Review Letters},
  volume = {48},
  number = {9},
  pages = {596--599},
  issn = {0031-9007},
  doi = {10.1103/PhysRevLett.48.596},
  urldate = {2025-02-14},
  copyright = {http://link.aps.org/licenses/aps-default-license},
  langid = {english}
}

@inproceedings{pucicDiffusionCopperGold1993,
  title = {Diffusion of Copper into Gold Plating},
  booktitle = {1993 {{IEEE Instrumentation}} and {{Measurement Technology Conference}}},
  author = {Pucic, S.P.},
  year = 1993,
  pages = {114--117},
  publisher = {IEEE},
  address = {Irvine, CA, USA},
  doi = {10.1109/IMTC.1993.382669},
  urldate = {2025-08-29},
  isbn = {978-0-7803-1229-6},
  langid = {english}
}

@article{raabTrappingNeutralSodium1987,
  title = {Trapping of {{Neutral Sodium Atoms}} with {{Radiation Pressure}}},
  author = {Raab, E. L. and Prentiss, M. and Cable, Alex and Chu, Steven and Pritchard, D. E.},
  year = 1987,
  month = dec,
  journal = {Physical Review Letters},
  volume = {59},
  number = {23},
  pages = {2631--2634},
  issn = {0031-9007},
  doi = {10.1103/PhysRevLett.59.2631},
  urldate = {2025-09-02},
  copyright = {http://link.aps.org/licenses/aps-default-license},
  langid = {english}
}

@article{riceAtmosphericCorrosionCopper1981,
  title = {Atmospheric {{Corrosion}} of {{Copper}} and {{Silver}}},
  author = {Rice, D. W. and Peterson, P. and Rigby, E. B. and Phipps, P. B. P. and Cappell, R. J. and Tremoureux, R.},
  year = 1981,
  month = feb,
  journal = {Journal of The Electrochemical Society},
  volume = {128},
  number = {2},
  pages = {275},
  publisher = {IOP Publishing},
  issn = {1945-7111},
  doi = {10.1149/1.2127403},
  urldate = {2025-09-16},
  langid = {english}
}

@article{sadiku-agboolaInfluenceOperationParameters2011,
  title = {Influence of {{Operation Parameters}} on {{Metal Deposition}} in {{Bright Nickel-plating Process}}:},
  shorttitle = {Influence of {{Operation Parameters}} on {{Metal Deposition}} in {{Bright Nickel-plating Process}}},
  author = {{Sadiku-Agboola}, O and Sadiku, E R and Ojo, O I and Akanji, O L and Biotidara, O F},
  year = 2011,
  journal = {Portugaliae Electrochimica Acta},
  volume = {29},
  number = {2},
  pages = {91--100},
  issn = {1647-1571},
  doi = {10.4152/pea.201102091},
  urldate = {2026-07-01},
  langid = {english}
}

@article{strehleElectricalPropertiesElectroplated2009,
  title = {Electrical Properties of Electroplated {{Cu}}({{Ag}}) Thin Films},
  author = {Strehle, S. and Bartha, J. W. and Wetzig, K.},
  year = 2009,
  month = apr,
  journal = {Thin Solid Films},
  volume = {517},
  number = {11},
  pages = {3320--3325},
  issn = {0040-6090},
  doi = {10.1016/j.tsf.2008.11.146},
  urldate = {2025-09-16}
}

@article{wangLongitudinalZeemanSlower2015,
  title = {A {{Longitudinal Zeeman Slower Based}} on {{Ring-Shaped Permanent Magnets}} for a {{Strontium Optical Lattice Clock}}},
  author = {Wang, Qiang and Lin, Yi-Ge and Gao, Fang-Lin and Li, Ye and Lin, Bai-Ke and Meng, Fei and Zang, Er-Jun and Li, Tian-Chu and Fang, Zhan-Jun},
  year = 2015,
  month = oct,
  journal = {Chinese Physics Letters},
  volume = {32},
  number = {10},
  pages = {100701},
  issn = {0256-307X, 1741-3540},
  doi = {10.1088/0256-307X/32/10/100701},
  urldate = {2025-02-17},
  copyright = {http://iopscience.iop.org/info/page/text-and-data-mining},
  langid = {english}
}

@article{wodeyRobustHighfluxSource2021a,
  title = {A Robust, High-Flux Source of Laser-Cooled Ytterbium Atoms},
  author = {Wodey, E and Rengelink, R J and Meiners, C and Rasel, E M and Schlippert, D},
  year = 2021,
  month = feb,
  journal = {Journal of Physics B: Atomic, Molecular and Optical Physics},
  volume = {54},
  number = {3},
  pages = {035301},
  issn = {0953-4075, 1361-6455},
  doi = {10.1088/1361-6455/abd2d1},
  urldate = {2025-08-20},
  langid = {english}
}

@article{yuZeemanSlowingGroupIII2022,
  title = {Zeeman Slowing of a Group-{{III}} Atom},
  author = {Yu, Xianquan and Mo, Jinchao and Lu, Tiangao and Tan, Ting You and Nicholson, Travis L.},
  year = 2022,
  month = mar,
  journal = {Physical Review Research},
  volume = {4},
  number = {1},
  pages = {013238},
  issn = {2643-1564},
  doi = {10.1103/PhysRevResearch.4.013238},
  urldate = {2025-02-14},
  langid = {english}
}

\end{document}